\documentclass[iop]{emulateapj}

\newcommand{\kms}{km s$^{-1}$}
\newcommand{\msun}{$M_{\sun}$}
\newcommand{\hubu}{km s$^{-1}$ Mpc$^{-1}$}

\newcommand{\ha}{H$\alpha$}
\newcommand{\hb}{H$\beta$}
\newcommand{\pb}{Pa$\beta$}
\newcommand{\bg}{Br$\gamma$}
\newcommand{\oiii}{{\rm [O~III]}}
\newcommand{\nh}{$N_{\rm H}$}
\newcommand{\av}{$A_{\rm V}$}
\newcommand{\mbh}{$M_{\rm BH}$}
\newcommand{\lbol}{$L_{\rm bol}$}
\newcommand{\ledd}{$L_{\rm Edd}$}
\def\wave#1{$\lambda${#1}}

\slugcomment{Received 2010 November 12; accepted 2010 December 06; published 2010 December 17}
\journalinfo{The Astrophysical Journal Letters: 726:L21, 2011 January 10}

\shortauthors{TRAN, LYKE, \& MADER}
\shorttitle{NAKED SEYFERT 2 GALAXIES}

\begin{document}

\title{Indecent Exposure in Seyfert 2 Galaxies: A Close Look}

\author{Hien D. Tran, J. E. Lyke, and Jeff A. Mader}
\affil{W. M. Keck Observatory, 65-1120 Mamalahoa Hwy., Kamuela, HI 96743, USA}
\email{htran@keck.hawaii.edu}

\begin{abstract}
NGC 3147, NGC 4698, and 1ES 1927+654 are active galaxies that are classified as 
Seyfert 2s, based on the line ratios of strong narrow emission lines 
in their optical spectra. However, they exhibit rapid X-ray spectral variability 
and/or little indication of obscuration in X-ray spectral fitting, contrary to 
expectation from the active galactic nucleus (AGN) unification model. Using optical spectropolarimetry 
with LRIS and near-infrared spectroscopy with NIRSPEC at the W. M. Keck Observatory, we 
conducted a deep search for hidden polarized broad \ha~and direct broad \pb~or 
\bg~emission lines in these objects. We found no evidence for any broad emission lines 
from the active nucleus of these galaxies, suggesting that they are unobscured, completely ``naked'' AGNs that 
intrinsically lack broad-line regions. 
\end{abstract}

\keywords{galaxies: active --- galaxies: individual (NGC 3147, NGC 4698, 1ES 1927+654) --- galaxies: Seyfert --- polarization}

\section{Introduction} \label{intro}

The ``unification model'' (UM) of Seyfert galaxies proposes that Seyfert 1 (S1) and Seyfert 2 (S2)
galaxies are basically the same type of objects viewed from different directions \citep{a93}.
In an S1 nucleus, our view of the central engine is relatively unobstructed, allowing a direct
observation of the ionizing continuum and broad-line region (BLR). In an S2 nucleus, however, this 
view is blocked by some form of obscuration, most often thought of as an optically thick torus, 
so that any broad emission lines (BELs) and ionizing continuum are not directly visible. While this UM has 
enjoyed considerable success over the years, and is undoubtedly correct for many Seyfert 
galaxies, questions regarding its universal applicability remain \citep[see, e.g.,][]{wz07}. Of 
particular interest is if there exists a ``true'' type of S2 that lacks a BLR, and whose 
appearance is unchanged regardless of orientation. The existence of such true or ``naked'' S2s has been 
implicated both observationally \citep{t01,pb02,t03,b03,nmm03,haw04} and theoretically 
\citep{nic00,l03,cze04,es06,eh09,cao10}. It would be of great interest to show definitively that such objects do 
exist in nature. Recently, a growing number of active galactic nuclei (AGNs), including NGC 3147, have been suggested to be just such objects \citep{gsf07,bi08,pan09,shi10}.
 
In this Letter, we explore three of the best candidates for such type of naked S2s: NGC 3147 
\citep{pap01,tw03,bi08}, NGC 4698 \citep{pap01,gz03}, and 1ES 1927+654 \citep{b03}. X-ray observations with {\it ASCA}, 
{\it ROSAT}, {\it XMM-Newton}, and {\it Chandra} show that all three exhibit characteristics typical of a type-1 view 
of the active nucleus: little or no intrinsic X-ray absorption from spectral fitting, high hard 
X-ray to \oiii~ratios indicating low obscuration \citep{b99}, and in the case of 1ES 1927+654, rapid, 
persistent, and strong X-ray variability observed over a 12 yr timescale. In addition, they are 
all classified as Compton thin, consistent with little or no intrinsic absorption above Galactic 
column density ($\sim$10$^{20}-10^{21}$ cm$^{-2}$). Thus, all indications appear to show that 
we have an unobscured, direct view of these active nuclei. Yet, optically they are classified as 
low-luminosity S2s, with no discernible sign of broad Balmer emission lines in their spectra \citep{ho97,b03}.
Table \ref{oxp} summarizes the observed and derived optical and X-ray properties of the galaxies. 
Given the type-1 view inferred from the X-ray observations, the lack of BELs in these objects is puzzling 
and in apparent contradiction with the UM of AGNs. We have therefore conducted a deep search for any 
weak or hidden BELs in these objects using optical spectropolarimetry and near-infrared (near-IR) spectroscopy. 
Deep spectropolarimetry was designed to target \ha~to look for any scattered, 
polarized broad-line component, or hidden broad-line region (HBLR), and near-IR spectroscopy was 
used to probe deeper through any obscuration to directly uncover any broad permitted lines, such 
as \pb~and \bg, as has been done by, e.g., \citet{vgh97}. This Letter reports the main results of this search. 
 
 \begin{deluxetable*}{lccccccccc}
\tablecolumns{10}
\tabletypesize{\scriptsize}
\tablewidth{0pt}
\tablecaption{Optical and X-ray Characteristics$^{a}$ \label{oxp}}
\tablehead{
\colhead{Object} & \colhead{Type} & \colhead{$z$} & \colhead{$m_{B}$} & \colhead{$f^{b}_{\rm [O~III]}$} & \ha/\hb & \colhead{\nh} & \colhead{$f_{2-10~keV}/f_{[\rm O~III]}$} & log(\mbh)  & log(\lbol/\ledd)  \\
\colhead{} & \colhead{} & \colhead{} & \colhead{} & \colhead{(10$^{-15}$ erg cm$^{-2}$ s$^{-1}$)} & \colhead{} & \colhead{(cm$^{-2}$)} & \colhead{} & \colhead{(\msun)} & \colhead{}
}
\startdata
NGC 3147 & Sey 2 & 0.0094   & 11.4 &  17.2$^{c}$ & 5.23$^{c}$& 1.5 $\times~10^{21}$ & $\sim$40 &  8.64  & ~~$ -3.05$ \\
NGC 4698 & Sey 2 & 0.0034   & 11.5 &  11.2~ & 3.12~  & 5 $\times~10^{20}$ & 1--3 &  7.43 & ~~$ -3.39$ \\
1ES 1927+654 & Sey 2 & 0.019~ & 15.4 &  3.16 & 4.15$^{d}$ & 7.3 $\times~10^{20}$ & $\approx 800^{e}$ & 7.34 & $\sim$$-2.23~$ \\ 
\enddata
\tablenotetext{a}{X-ray data are from \citet{tw03}, \citet{gz03}, and \citet{b03}. Unless otherwise noted, optical data are from this study. A cosmology of $H_o$ = 70 \hubu, $q_o$ = 0.5, and $\Lambda$ = 0 is assumed.}
\tablenotetext{b}{Observed \oiii~\wave 5007 flux uncorrected for extinction measured from galaxy-subtracted spectrum.}
\tablenotetext{c}{From \citet{ho97}.}
\tablenotetext{d}{Due to the strong absorption at \hb~in the nuclear spectrum, only an upper limit can be determined \citep{b03}. This value is measured from our extended spectrum just off the nucleus, free of \hb~absorption.}
\tablenotetext{e}{Only a 0.3-7 keV flux is available \citep{b03}. Using the photon index $\Gamma = 2.7$, we estimate the hard X-ray (2-10 keV) flux,  $f_{2-10~keV}$  $\sim$ 0.2$f_{0.3-7~keV}$.}
\end{deluxetable*}
 
\section{Observations and Data Reduction} \label{obs}

Spectropolarimetric observations were made with the low-resolution imaging spectrograph 
\citep[LRIS;][]{oke95} and polarimeter on the Keck I telescope at the W. M. Keck Observatory (WMKO).
We used a long, 1\arcsec~wide slit centered on the nucleus of the AGN. For NGC 4698 and 1ES 1927+654,
multiple observations were made similar to that described in \citet{t10} over two different epochs to improve the 
signal-to-noise ratio (S/N) and to look for any evidence of variability.  No variability was detected, and 
the results presented here are the sum average for all observations over all epochs for each object.
The total exposure times in Table \ref{olog} represent some of the deepest spectropolarimetric 
observations for this type of objects, typically $\sim$10$\times$ deeper than any previous surveys
\citep[see, e.g.,][]{t03}. Spectropolarimetric reduction was done with standard techniques, as 
described in \citet{t95}. 

\begin{deluxetable*}{lcccccc}
\tablecolumns{7}
\tabletypesize{\scriptsize}
\tablewidth{0pt}
\tablecaption{Journal of Observations \label{olog}}
\tablehead{
\colhead{Object} & \colhead{UT Date} & \colhead{Instrument} & \colhead{Total Exposure} & \colhead{Slit} & \colhead{Filter} & \colhead{$\lambda$ Range}  \\
\colhead{} & \colhead{} & \colhead{} &\colhead{(s)} & \colhead{} & \colhead{} & \colhead{} 
}
\startdata
NGC 3147 & 2003 Jun 28 & LRISp   & 4$\times$900 & 1\arcsec & \nodata & 6000 -- 9800 \AA \\
         & 2010 Feb 25 & NIRSPEC & 3600  & 42\arcsec $\times$ 0\farcs76 & N-3 & 1.15 -- 1.35 $\mu$m \\
         & 2010 Feb 25 & NIRSPEC & 1600  & 42\arcsec $\times$ 0\farcs76 & K & 2.00 -- 2.43 $\mu$m \\
NGC 4698 & 2003 Jun 28 & LRISp   & 2$\times$(4$\times$900) & 1\arcsec & \nodata & 6000 -- 9800 \AA \\
         & 2004 Jun 17 & LRISp   & 4$\times$1200 & 1\arcsec & \nodata & 4650 -- 7200 \AA \\
         & 2005 May 24 & NIRSPEC & 2880  & 42\arcsec $\times$ 0\farcs76 & N-3 & 1.15 -- 1.35 $\mu$m \\
         & 2010 Feb 25 & NIRSPEC & 2400  & 42\arcsec $\times$ 0\farcs76 & K & 2.00 -- 2.43 $\mu$m \\
1ES 1927+654 & 2003 Jun 28 & LRISp & 2$\times$(4$\times$1200) & 1\arcsec & \nodata & 6000 -- 9800 \AA \\ 
             & 2004 Jun 17 & LRISp & 4$\times$1200 & 1\arcsec & \nodata & 4650 -- 7200 \AA \\
             & 2004 May 31 & NIRSPEC & 1920 & 42\arcsec $\times$ 0\farcs57 & N-3 & 1.15 -- 1.35 $\mu$m \\
             & 2004 Jul 21 & NIRSPEC & 2880 & 42\arcsec $\times$ 0\farcs57 & K & 2.00 -- 2.40 $\mu$m \\
\enddata
\end{deluxetable*}

Near-infrared spectroscopic observations were made with NIRSPEC \citep{mcl98} on the Keck II 
telescope at WMKO in the low-resolution mode with the 42\arcsec~$\times$ 0\farcs57 or 0\farcs76 slit. 
The telescope was dithered in an ABBA pattern, and nearby A0 V stars (39 UMa, 55 Dra, HD 111744, and
HIP 62745) were observed to correct for telluric absorption and relative flux calibration. Data reduction was initially 
performed with the REDSPEC\footnote{http://www2.keck.hawaii.edu/inst/nirspec/redspec.html} package 
for flat fielding, spectral rectification, wavelength calibration, and spectral extraction. Exposures from 
different nod positions were used for sky subtraction and then co-added. Individual exposures range from 240 s 
to 900 s and the total integration time for each object is shown in Table \ref{olog}. We then used the routine 
{\it xtellcor\_general} \citep{vac03} within the Spextool package \citep{cus04} to perform telluric correction 
and relative flux calibration. Table \ref{olog} presents the log of observations.  

\section{Results and Discussion} \label{disc}

\begin{figure}
\includegraphics[scale=0.35,angle=-90]{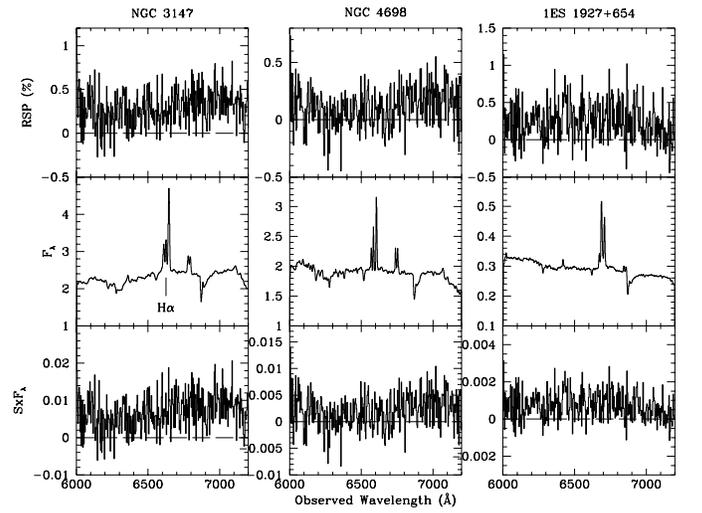}
\caption{Spectropolarimetry of NGC 3147, NGC 4698, and 1ES 1927+654. 
Top: rotated Stokes parameter (RSP), essentially a measure of the polarization in percent; 
middle: total flux spectrum ($F_\lambda$) ; bottom: the polarized or Stokes flux spectrum ($S\times F_\lambda$).
The flux scales are in units of 10$^{-15}$ erg s$^{-1}$ cm$^{-2}$ A$^{-1}$.
The position of \ha~is marked. In each object, a small amount of polarization is detected but no 
polarized broad lines indicative of an HBLR are seen in the polarized flux spectra. 
\label{spol}}
\end{figure}

\begin{figure}[b]
\includegraphics[scale=0.35,angle=-90]{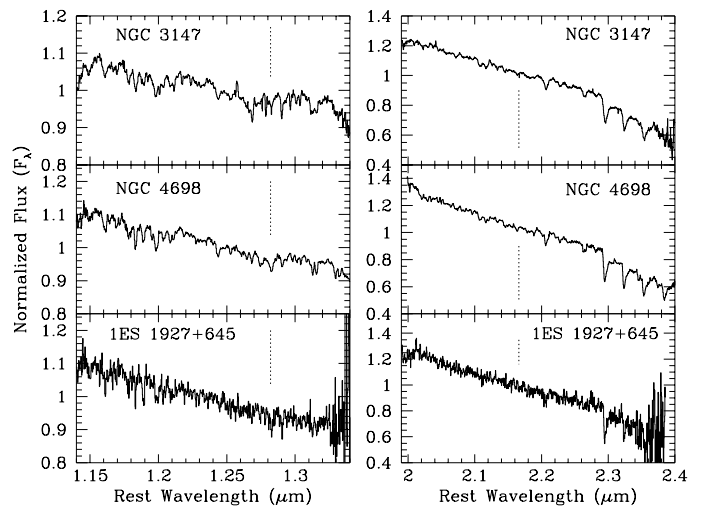}
\caption{Normalized $J$- and $K$-band spectra of NGC 3147, NGC 4698, and 1ES 1927+654. In each case, the expected 
positions of \pb~(left) and \bg~(right) are marked as dotted vertical lines. The spectra are dominated by galactic starlight, and we do not detect any emission in \pb~or \bg. No direct BELs are present. 
\label{nspec}}
\end{figure}

We show in Figure \ref{spol} the results of the optical spectropolarimetry, and in Figure \ref{nspec} those of the near-IR spectroscopy. Table \ref{spres} summarizes the polarimetric characteristics of the objects. The main spectropolarimetric result is that in each of the objects, although a small polarization is detected, no polarized broad \ha~is seen. 
Note that the polarizations listed in the table are the observed values, uncorrected for galactic host starlight dilution. With typical galaxy fractions of $\sim$90\%, the intrinsic polarizations are $\sim$1\%--3\%.
The small measured polarizations appear intrinsic to each galaxy, as the Galactic interstellar polarizations are probably insignificant due to their relatively high Galactic latitudes (Table \ref{spres}). Moreover, no narrow emission lines are visible in the polarized flux spectra, indicating that any polarization imposed outside of the host galaxies is negligible.  They belong to the class of non-HBLR S2s \citep{t03}.  The non-detection of HBLRs is secure, confirming similar results of \citet{shi10} for NGC 3147 and NGC 4698. HBLRs have been successfully detected in sources with comparable weakly polarized continuum such as NGC 2110 \citep{t10}.

Likewise,  although the S/N is excellent in the near-IR spectroscopy, there is no detection of any of the broad permitted
hydrogen lines such as \pb~or \bg. The near-IR spectra are dominated by the underlying host stellar continuum, with strong absorption band heads due to CO, but essentially no emission lines are detected. This is very similar to the near-IR spectra of some objects in the atlas of \citet{rif06}, such as NGC 1144 and NGC 1097. 

We now examine why we do not see any BELs given the naked nature of these AGNs, which we shall refer to as the ``naked non-HBLR S2s'' (NNHS2s). The possible explanations include: (1) highly obscured AGNs that are misclassified as Compton thin, (2) different obscuration in the X-ray and optical (objects may be X-ray-unobscured but are actually highly obscured optically), (3) variable BELs due to either changes in the nuclear engine or to obscuring material moving in and out of our line of sight, (4) hidden narrow-line S1 galaxies (NLS1s), and (5) low-powered AGNs with weak or absence of BLRs. We discuss below each possibility in turn. 

\begin{deluxetable}{lcccc}[t]
\tablecolumns{3}
\tabletypesize{\scriptsize}
\tablewidth{0pt}
\tablecaption{Spectropolarimetric Results \label{spres}}
\tablehead{
\colhead{Object} & \colhead{$b_{\rm II}^a$} & \colhead{$E(B-V)^b$} &  \colhead{$p^{c}$} & \colhead{$\theta^{c}$}  \\
\colhead{} &  \colhead{(\arcdeg)} & \colhead{$p_{\rm max}$} & \colhead{(\%)} & \colhead{(\arcdeg)} 
}
\startdata
NGC 3147 & 39.5 & 0.024 & 0.31 $\pm$ 0.01 &  159  $\pm$ 1  \\
                    &          & 0.22\% &                                &    \\
NGC 4698 &  71.3 & 0.026 & 0.11 $\pm$ 0.01 &  ~64  $\pm$ 3  \\
                    &          & 0.23\% &                                &    \\
1ES 1927+654 & 21.0  & 0.088 & 0.26 $\pm$ 0.02 &  ~88  $\pm$ 2  \\ 
                    &          & 0.79\% &                                &    \\
\enddata
\tablenotetext{a}{Galactic latitude.}
\tablenotetext{b}{Galactic interstellar reddening from NED. The maximum Galactic interstellar polarization, $p_{\rm max}$ obeys the relation $p_{\rm max} \leq 9E(B-V)$ \citep{ser75}.}
 \tablenotetext{c}{Observed average over the wavelength range 6050--7100\AA.} 
\end{deluxetable}

1. All three NNHS2s in this Letter are classified as Compton thin due to their low column densities \nh~$\lesssim 10^{21}$ cm$^{-2}$, derived from X-ray spectral fitting. In addition, the hard X-ray to \oiii~flux ratio $f_{2-10keV}/f_{[\rm O~III]}$, a good measure of the nuclear X-ray obscuration \citep{b99}, is comfortably above unity, the usual dividing line between Compton-thin and Compton-thick cases (see Table \ref{oxp}). This is similar to the ratio in S1s, where it is typically $\sim$20 with a range between about 1 and 300 \citep{b99,ag09}, again consistent with little intrinsic absorption. 

Recently \citet{shu10}, using higher spatial resolution {\it XMM-Newton} observations, find that the X-ray emission from the naked S2 candidate NGC 7590 is actually dominated by extended off-nuclear sources, leading them to conclude that this galaxy is actually Compton thick rather than X-ray unobscured as previously thought. If a similar situation applied to the NNHS2s in this study, then they could also be misclassified as X-ray unobscured. However, all three objects, NGC 3147, NGC 4698, and 1ES 1927+654, have been extensively studied with high-resolution {\it Chandra} and {\it XMM-Newton} observations \citep[e.g.,][]{b03,gz03,tw03,bi08,bn08,ag09,shi10}, and the unabsorbed nature of their nuclear X-ray emission appears to be well confirmed, with no confusion from external sources. In addition, temporal variation in the X-ray flux has been observed in both NGC 3147 \citep{tw03} and 1ES 1927+654 \citep{b03}, implying that the X-ray flux is not scattered from a heavily obscured nucleus. Thus, all three objects in this study appear to be genuinely naked S2s. 

2. Studies have shown that the X-ray and optical nuclear absorption in AGNs are often mismatched \citep[see review by][]{mr07}. The reason may be that the X-ray and optical absorbers are not cospatial. Some Seyfert galaxies are known to undergo extreme changes in X-ray column density \nh, perhaps due to the gaseous absorbing material crossing the line of sight to the X-ray source, as seen in, e.g., NGC 1365 \citep{ris07}. The observed rapid \nh~variability has constrained the X-ray absorber location to the scale of the BLR.  The size scale of the nuclear optical absorber, however, is probably much larger than the dust sublimation radius or beyond the BLR. As a result, it is common to find, for example, Compton-thick AGNs that are seemingly optically unobscured, with type-1 BELs \citep{mr07}.  

Could the objects in this study be the Compton thin but optically thick AGNs? This could be similar to the class of X-ray bright optically normal galaxies (XBONGs), which are strong in X-ray, but optically dull, leading to the suggestion that they may be optically obscured by large-scale dust in the host galaxy \citep{rig06}. 
For the objects in this study the extinction inferred from the X-ray column density is only \av~$\sim 0.2$$-$0.68 assuming the Galactic relationship between \nh~and \av~of \citet{g75}.
If  they were obscured by galactic-scale dust similar to XBONGs, they may be expected to show much higher extinction in the narrow-line region (NLR).  However, as Table \ref{oxp} shows, the narrow-line Balmer decrements are fairly normal, yielding at most an optical extinction of \av~$\le 1.6$, indicating that heavy large-scale dust obscuration cannot explain the lack of BLR in these galaxies. If typical broad permitted lines were present, their complete absence in both the optical and near-IR implies that the extinction must be of order \av~$\sim 11$$-$26 \citep{vgh97}, entirely inconsistent with observations.  While it is possible that the gas to dust ratio \nh/\av~could be anomalously low in these galaxies, leading to a preferentially higher absorption in the optical, it cannot explain this discrepancy. Although we cannot rule out such heavy obscuration in the BLR itself, it seems unlikely.  

The absence of any narrow lines in the polarized flux spectrum also indicates that the scattering region must be interior to the NLR\footnote{We assume, as in other Seyfert galaxies, that the detected polarizations come from scattering of light by an anisotropic medium such as a bi-cone.}. The scattering region may be very compact, as in the hidden double-peaked emitters (HDPEs) NGC 2110 and NGC 5252, lying between the NLR and BLR \citep{t10}, such that it could still be obscured from the line of sight. However, if this were the case, we would not expect to see any polarized light. The fact that we see any scattered light at all suggests that the scattering region, however small, is intercepted, and that the lack of any BELs suggests that there may be very little or no BLR gas, or that the scatterers are within the BLR. 
 
3. Changes in the central engine or the obscuration could result in changes in appearance of the AGN. For example, NGC 2110 and NGC 5252 were originally classified as non-HBLR S2s, but perhaps due to their intrinsic broad-line variability, later found to be HDPEs \citep{t10}.
Could the objects in this study be similar to such objects? This is unlikely since multiple observations intended to look for such variability failed to find any. For NGC 3147 and NGC 4698, there is an additional spectropolarimetric epoch reported by \citet{shi10}, who also failed to detect any HBLRs. While more follow-up observations may be required to definitively rule out variability, the present data do not favor such scenario. 

Another cause of changes in classification would be for optically obscuring clouds to pass in front of the line of sight, changing the appearance of the objects \citep[e.g.,][]{tom92}. This type of changes typically occurs over very long timescales, of order years or decades. Since all available published data over many years show no evidence for the appearance or disappearance of BELs in any of the objects, we do not favor this interpretation for the lack of HBLR in these NNHS2s. 

4. Could the NNHS2s be the hidden counterparts to the NLS1s, as suggested by, e.g., \citet{dg05}, \citet{zw06}, \citet{wz07}, 
and \citet{has07}? This can be ruled out by the simple observation that no emission lines of any kind, broad or narrow, are seen in the polarized flux spectra. In addition, one defining characteristics of NLS1s is the strong Fe II emission \citep{op85}, but none is observed in the polarized light of these NNHS2s. Polarized Fe II emission has been detected in other HBLR S2s \citep{t99}.
To be sure, the hidden NLS1s envisioned by  \citet{zw06} and \citet{wz07} are thought to be those with high X-ray absorption, not the X-ray unobscured type discussed here. 

5. The most likely scenario that can explain the lack of BELs is that NNHS2s represent true S2s with very little or virtually no BLR. These are the low-luminosity AGNs, probably powered by radiatively inefficient or advection dominated accretion flow (ADAF), that intrinsically lack BLRs, as suggested observationally by, e.g., \citet{t01,t03}, \citet{bi08}, \citet{pan09}, and 
\citet{shi10}, and inspired theoretically by \citet{nic00}, \citet{l03}, \citet{es06}, \citet{eh09}, and \citet{cao10}. They could be at the dead and dying endpoint in the evolutionary path envisioned by Wang \& Zhang (2007; their ``unabsorbed non-HBLR S2s B'' designation). As suggested by \citet{wz07}, these could be the gas-poor, and dust-rich galaxies, which after undergoing vigorous star formation and accreting matter at a high rate have exhausted all their fuel and torus material, and are now lying dormant and naked. Note that this would naturally lead to the low gas-to-dust ratio discussed above. 

Since there is a well-known luminosity$-$BLR size relationship in AGNs \citep[see, e.g.,][]{den10}, the low luminosities of the NNHS2s may lead to BLRs that are exceptionally small, resulting in lines that are too wide to be detectable \citep{l03}. We note, however, that it is possible to detect hidden extremely broad lines (FWHM up to $\sim$26,000 \kms) in objects with brightness and continuum polarization levels comparable to those in this study \citep{o97,t10}.   
Another key point in the theoretical models mentioned above is that the BLR is unable to form once the AGN luminosities or accretion rates become too low to support outflows from the accretion disk. The critical accretion rate at which the BLR is expected to disappear is \lbol/\ledd~$\sim$ 0.001 \citep[e.g.,][]{nmm03,cao10}, with a weak dependence on the black hole (BH) mass\footnote{In the \citet{nic00} models, the critical \lbol/\ledd~$\sim$ 0.024(\mbh/\msun)$^{-1/8}$, ranging from $\sim$0.0043 for a $10^{6}$ \msun~BH to $\sim$0.0013 for a $10^{10}$ \msun~BH.}.
We use the measured \oiii~FWHM (403 \kms) and luminosity to estimate the BH mass \mbh~and accretion rate \lbol/\ledd~for 1ES 1927+654, using the method outlined in \citet{wz07}. These quantities are listed in Table \ref{oxp}~along with those for NGC 3147 and NGC 4698, which are obtained from the literature \citep[e.g.,][]{wz07}.  Given the uncertainties in estimating \mbh~and \lbol, the uncertainties in log(\lbol/\ledd)~is $\sim$ 0.5 dex. As the table shows, all of the accretion rates are consistent with being below the minimum threshold needed to support BLRs. Thus, the lack of BLRs in these NNHS2s can be entirely attributed to the feebleness of their central engines. 

Similarly, a significant fraction of the XBONGs can also be explained by such lethargy in their nuclear activity \citep{tru09}. The NNHS2s, especially 1ES 1927+654 because of its high X-ray-to-optical flux ratio, could very well be the local population of these XBONGs. 
In addition, there is evidence that many Fanaroff$-$Riley I radio galaxies are also missing hidden BLRs \citep[see review by][]{a11}, and thus they could potentially be the corresponding low accretion-powered radio-loud population of NNHS2s.
We note that there is also another class of AGNs with ``anemic'' BLR \citep{she10}, but these appear to be different from the NNHS2s discussed here since they are powerful quasars at higher redshifts with much higher accretion rates. 

\section{Summary and Conclusions}

NGC 3147, NGC 4698, and 1ES 1927+654 are three S2s with an unusual combination of 
properties: X-ray spectra show variability and little absorption indicative of a type-1 (direct) 
view, but optical spectra show only narrow emission lines, typical of a type-2 (obscured) view of 
the nucleus. A deep search for hidden BLR using Keck LRIS spectropolarimetry and direct near-IR 
spectroscopy with NIRSPEC does not reveal any BELs, hidden or direct. If typical broad lines 
were present, their non-detections would indicate an extinction of \av~$\sim$ 11$-$26, inconsistent
with the ``naked'' nature of these galaxies.  While the obscuration may be due to different material for X-ray and 
optical light, it appears plausible that the BLRs in these objects are anemically small or absent, due to the weakness 
of their active central engines. 

\acknowledgments
We thank the WMKO directors, F. Chaffee and T. Armandroff, for making
this research possible with access to their director's time through Team Keck. 
We thank the referee for constructive comments.
The data presented herein were obtained at WMKO, which is operated as a scientific partnership among the California Institute of Technology, the University of California, and the National Aeronautics 
and Space Administration (NASA). The Observatory was made possible by the generous financial support of 
the W. M. Keck Foundation. 
This research has made use of NED and the Keck Observatory Archive (KOA), which is operated by WMKO and the NASA Exoplanet Science Institute, under contract with NASA. 

{\it Facilities:} \facility{Keck I (LRISp)}, \facility{Keck II (NIRSPEC)}, \facility{KOA}
 


\end{document}